\documentclass{JHEP3}
\usepackage{epsfig}
\usepackage{amssymb,latexsym}

\newcommand{\beq}{\begin{equation}}
\newcommand{\eeq}{\end{equation}}
\newcommand{\ba}{\begin{array}}
\newcommand{\ea}{\end{array}}
\newcommand{\bea}{\begin{eqnarray}}
\newcommand{\eea}{\end{eqnarray}}

\def\p{\partial}
\def\pb{{\bar \partial}}
\def\u{\upsilon}
\def\ub{{\bar \upsilon}}

\preprint{MIT-CTP-3246\\  {\tt hep-th/0202129}}

\title{The partition function of the two-dimensional black hole conformal
field theory}
\author{Amihay Hanany, Nikolaos Prezas and Jan Troost
\\
Center for Theoretical Physics,
\\ Massachusetts Institute of Technology,\\ Cambridge, MA 02139, USA.\\
\email{hanany, prezas, troost@mit.edu}
}
\abstract{We compute the partition function of the conformal field theory
on the two-dimensional
euclidean black hole background using path-integral techniques. We show that
the resulting spectrum is  consistent with the algebraic expectations
for the $SL(2,R)/U(1)$ coset conformal field theory
construction. In particular, we find confirmation for the bound on the
spin of the discrete representations and we determine
the density of the continuous
representations. We point out the relevance of the partition function
to all string theory backgrounds that include an $SL(2,R)/U(1)$ coset
factor. }
\begin{document}
\section{Introduction}
Recently, substantial progress was made in understanding non-compact
Wess-Zumino-Witten models. In particular, the spectrum and correlation
functions of string theory on $AdS_3$, the covering space of $SL(2,R)$,
was analysed in detail in 
\cite{Maldacena:2000hw, Maldacena:2000kv, Maldacena:2001km}. The 
spectrum for the non-compact Wess-Zumino-Witten model
was determined in
\cite{Maldacena:2000hw}, using intuition for long strings obtained from
\cite{Maldacena:1998uz, Seiberg:1999xz}
and the technical
tool of spectral flow \cite{Schwimmer:mf}, thereby solving the
long-standing problem of determining the correct Hilbert space for
the $SL(2,R)$ WZW-model
\cite{Balog:1988jb,Petropoulos:1989fc,Mohammedi:1989dp,Evans:1998qu,Bars:1990rb,Hwang:1990aq,Hwang:1991an,Hwang:1998tr,Henningson:1991jc,Hwang:1992uk,Kato:2000tb,Dixon:1989cg}.
Next, in \cite{Maldacena:2000kv}, the computation
of the free energy for string theory on $AdS_3$\footnote{See also e.g.
\cite{Giveon:1998ns,Kutasov:1999xu,deBoer:1998pp}.}
by path-integral methods
gave additional support to the spectrum proposed in \cite{Maldacena:2000hw}.
Finally, in \cite{Maldacena:2001km}  the completeness of the Hilbert space
was checked by computing various correlators
(see also e.g.
\cite{Teschner:1997ft,Teschner:1997fv,Teschner:1999ug,Giribet:2001ft,Ishibashi:2000fn} 
for earlier
work).
This series
of papers has answered important questions in
non-compact Wess-Zumino-Witten theories and opened the road to a more
extensive study of these models.

It is natural then to re-address some old questions. In particular,
we can re-analyze \cite{Gawedzki:1991yu}
the toroidal
partition function for the $SL(2,R)/U(1)$ coset conformal field theory.
It is an important task to write down the partition function for this
background for several reasons.
The $SL(2,R)/U(1)$ background 
\cite{Bardakci:1987ee, Altschuler:1987zb, 
Rocek:1991vk,Mandal:1991tz,Elitzur:cb,Giveon:1991sy,Dijkgraaf:1991ba} was 
identified and  analysed as a
two-dimensional euclidean black hole in \cite{Witten:1991yr}. Subsequently,
it was employed in the construction of conformal field theories
describing exact string propagation on curved spacetimes (see e.g.
\cite{Antoniadis:1994sr}).
Moreover, many interesting string theory
backgrounds contain  $SL(2,R)/U(1)$ factors, for example
in the context
of singularities of Calabi-Yau manifolds \cite{Ooguri:1995wj},
holographic duals for NS5-brane backgrounds
\cite{Sfetsos:1998xd,Giveon:1999zm,Giveon:1999px}, et cetera.
In order to write down the toroidal
partition function for string theory on these backgrounds, the most crucial
ingredient is the black hole background partition sum. A precise treatment
of the spectrum and the Hilbert space for the coset model was hitherto 
lacking.

In this letter, we address the computation of the partition function
from a path-integral point of view. Our computation is technically
close to the analysis of the free energy of string theory on
$AdS_3$ in \cite{Maldacena:2000kv}.
In section \ref{setup} we discuss
the setup for our computation, discussing a few general features
of coset CFT and toroidal partition functions. In section \ref{computation}
we perform the actual computation and discuss the crucial ingredient
of Ray-Singer torsion.
Next, we analyse the result and show that it agrees with
the algebraic expectations in section \ref{analysis}. We conclude
and discuss applications in section \ref{conclusions}.

\section{The $SL(2,R)/U(1)$ coset toroidal partition function}
\label{setup} We introduce the model in this section and pay some
attention to the holonomies that will play a crucial role in the
computation of the partition function. The general treatment of
gauged Wess-Zumino-Witten models is well-known
\cite{Karabali:1988au,Karabali:1989dk,Gawedzki:1988hq,Gawedzki:1988nj}.
For a general group manifold $G$
the Wess-Zumino-Witten action is: \beq
S[g] = \frac{k}{2\pi} \int_{\rm WS} d^2z \;{\rm Tr}(\p g^{-1} \pb
g) + \frac{i k}{12\pi} \int_{\rm B} {\rm Tr} (g^{-1} dg)^3
\label{eq:wzw} \eeq where $g(z,{\bar z})$ is a group element,  the
level of the WZW-model is $k \in \mathbb{R}$, and $B$ is a
three-dimensional manifold with the worldsheet as a boundary. Our
worldsheet is a two-torus $T^2$. 

For compact group manifolds,
the level $k$ is in general quantised but for $SL(2,R)$ we have
$H^3(SL(2,R),\mathbb{R})=0$ so that the action is independent of our
choice of manifold $B$ for any real $k$.
The Wess-Zumino-Witten model has
an affine symmetry $G(z) \times G({\bar z})$. We will gauge an
axial abelian subgroup of the symmetry group $g \rightarrow
hgh$, yielding the action:
\beq S_{gauged}[g;A] = S[g] -\frac{k}{\pi} \int_{\rm WS} d^2 z \;{\rm Tr}
(\bar{A} \pb g g^{-1} + A g^{-1} \p g + g^{-1} \bar{A} g A + A
\bar{A}) \label{eq:gwzw} \eeq with the one-form gauge field
$A_{(1)}$ defined as $A_{(1)} = A \, dz + {\bar A} \, d{\bar z}$.
Our gauged theory is anomaly free (see e.g.
\cite{Ginsparg:1992af}).

Next we concentrate
on the Lorentzian $SL(2,R)$ group manifold. A standard parametrisation
of the group elements $g$ is in terms of Euler angles 
\cite{Dijkgraaf:1991ba}:
\begin{eqnarray}
g &=& e^{\frac{i}{2} \theta_L \sigma_2} e^{\frac{r}{2}  \sigma_1}
e^{\frac{i}{2}\theta_R \sigma_2 }
\end{eqnarray}
where $0 \leq r \leq \infty, 0 \leq \theta_L < 2 \pi , -2 \pi \leq
\theta_R < 2 \pi$. We gauge the axial U(1) symmetry $ g \rightarrow h g h$
where $h = \exp(\frac{i}{2}
\lambda \sigma_2)$. The coordinates shift under gauge
transformations as
$\theta_{L,R} \rightarrow \theta_{L,R} + \lambda$ and
the gauge field  transforms as
\beq
A \longrightarrow A + d\lambda.
\eeq
The gauged WZW action (\ref{eq:gwzw}) becomes:
\begin{eqnarray}
& S[r,\theta_R,\theta_L;A] =   \frac{k}{2 \pi} \int d^2 z
\Big(\frac{1}{2} (\p r \pb r - \p\theta_L\pb\theta_L -
\p\theta_R\pb\theta_R  - 2 \cosh r \pb\theta_L\p\theta_R) + & \nonumber \\
& (A (\pb\theta_R + \cosh r \pb \theta_L) + {\bar A}(\p\theta_L
+\cosh r \p\theta_R) - A{\bar A}(\cosh r + 1) \Big). &
\label{eq:gsl2}
\end{eqnarray}
In \cite{Witten:1991yr} it was shown, by integrating out the gauge
field classically, that the coset has a cigar geometry that can be
interpreted as a Euclidean black hole. Gauging a non-compact
abelian subgroup  would have resulted in the Lorentzian two-dimensional
black hole.

The gauged theory can be re-written in terms of the sum of an
$SL(2,R)$ model and a $U(1)$ model. We thereto introduce the
coordinates $\theta=\frac{1}{2} (\theta_L - \theta_R)$ and
$\tilde{\theta}= \frac{1}{2} (\theta_L + \theta_R)$. In terms of
these coordinates the action can be written in the manifestly
gauge invariant form (since shifts $\tilde{\theta} \rightarrow 
\tilde{\theta} + \lambda$
are compensated by shifts in the gauge field $A \rightarrow A + \partial
\lambda$ and $\bar{A} \rightarrow \bar{A} + \bar{\partial} \lambda$):
\begin{eqnarray}
& S[r,\theta,{\tilde \theta};A,{\bar A}]  =  \frac{k}{2 \pi} \int
d^2 z \Big[\frac{1}{2} \p r \pb r + (\cosh r - 1)
\Big(\p\theta\pb\theta + (A-\p{\tilde \theta})\pb\theta - ({\bar
A} - \pb{\tilde \theta})\p\theta\Big)
& \nonumber \\
& -(\cosh r + 1)(A-\p{\tilde \theta})({\bar A} - \pb{\tilde
\theta}) \Big].& \label{eq:gsl2-other}
\end{eqnarray}
To re-write the action in product form, we
first Hodge-decompose the gauge field on the torus as:
\begin{eqnarray}
A & = &  \p\rho_L + \frac{i}{2\tau_2}(u_1 {\bar \tau} - u_2)   \\
{\bar A} & = &  \pb\rho_R - \frac{i}{2\tau_2}(u_1\tau - u_2)
\end{eqnarray}
where $\rho_L^{\ast} = \rho_R$ is well-defined on the torus and
the holonomies $u_1, u_2$, parametrize the Wilson lines on the
toroidal worldsheet
with modular parameter $\tau$. Ignoring the holonomies for now, we
can follow the treatment on the sphere (as in e.g. \cite{Tseytlin:1992ri}).
We introduce the new variables
$\rho = \frac{1}{2} (\rho_L - \rho_R)$,
$\tilde{\rho} = \frac{1}{2} (\rho_L + \rho_R)$ and $\kappa=\theta+\rho$,
$\tilde{\kappa}=\tilde{\theta}-\tilde{\rho}$, in terms of which the
action becomes:
\begin{eqnarray}
& S[r,\kappa,{\tilde\kappa};\rho,{\tilde\rho}] = \frac{k}{2\pi}\int d^2 z
\Big(\frac{1}{2}\p r \pb r +
(\cosh r -1)\p\kappa\pb\kappa
-(\cosh r + 1)\p{\tilde \kappa}\pb{\tilde \kappa} + & \nonumber \\
& (\cosh r -1)(\p\kappa\pb{\tilde\kappa} - \pb\kappa\p{\tilde\kappa}) \Big)
+ \frac{k}{\pi} \int d^2 z \; \p\rho\pb\rho. &
\label{eq:sl2u1}
\end{eqnarray}
Since under a gauge transformation
$\rho_{L,R} \rightarrow \rho_{L,R} + \lambda$ the fields $\kappa,
{\tilde \kappa}$ and $\rho$ do not transform, the above action is
 gauge invariant.

We can read (\ref{eq:sl2u1}) as the action for
the $SL(2,R) \times U(1)$ model. Of course, on a toroidal worldsheet
we need to take care in following the holonomies in the gauge field
through the coordinate redefinitions. Note however that already for
the spherical topology, subtleties arise that have hitherto been
succesfully ignored.
Indeed, the quantity $\kappa$ is a linear combination of a real
and imaginary field, but will nevertheless be treated as a real
field \cite{Dijkgraaf:1991ba,Tseytlin:1992ri}. We will
briefly return to the subtle issues of analytic continuation in
the following, although we will not resolve all of them
unambigously. The above decomposition was put to good use in
\cite{Dijkgraaf:1991ba} to argue for the spectrum of the CFT and
to calculate the exact black hole background, and in
\cite{Tseytlin:1992ri} to compute the effective action and
rederive the exact metric in a path-integral approach. We note
that the holonomies of the gauge field have been transformed, via
the field redefinitions, into non-trivial windings (over the two
1-cycles of the torus) for the matter fields
$\kappa,{\tilde\kappa}$ and $\rho$. They will be crucial in the
following.\footnote{Our toroidal treatment of the holonomies
will naturally turn
out to be equivalent to the BRST analysis of the gauge invariant
states in \cite{Dijkgraaf:1991ba}.}

Before delving into the main part of the computation of the partition
function we  gauge-fix the action by choosing $\tilde{\rho}=0$.
We then need to include the ghost action
\beq
S_{ghosts}[b,c] = \frac{1}{\pi} \int d^2 z (b\pb c + {\bar b}\p {\bar c}).
\eeq
\section{Computing the partition function}
\label{computation}
This section contains the core of the computation of the toroidal
partition function. We will discuss the various techniques needed
for the computation in some detail. Our computation owes a lot
to the analysis in  \cite{Maldacena:2000kv,Gawedzki:1991yu}.
Of course, since \cite{Maldacena:2000kv} computes the free energy of $AdS_3$
string theory while we are interested in the partition function
on the euclidean black hole background, we need to adapt
their computational techniques creatively.\footnote{To avoid confusion, note
that the temperature
introduced in \cite{Maldacena:2000kv} is the temperature of $AdS_3$. The
euclidean black hole is an analytically continued version of the Lorentzian
black hole with a different time direction.}

Our previous treatment of the model was in accord with standard
conventions on Euler angles, but
to make the computation of the partition function feasible, it is
very useful to parametrize the $SL(2,R)$ part of the model in terms
of the coordinates introduced in \cite{Gawedzki:1991yu}.
After continuing the path integral to Euclidean signature to make it well
defined (effectively transforming the model into the $SL(2,C)/SU(2)$
coset model -- for discussions see \cite{Gawedzki:1991yu} and
\cite{Maldacena:2001km}), the coordinate transformation becomes:
\begin{eqnarray}
\u & = & \sinh \frac{r}{2} e^{i \kappa} \\
\ub & = & \sinh \frac{r}{2} e^{- i \kappa} \\
\phi & = & i {\tilde \kappa} - \log \cosh \frac{r}{2}
\label{eq:mos-us}.
\end{eqnarray}
Writing the total action in terms of these variables results in:
\begin{eqnarray}
S[\phi,\u,\ub;\rho;b,c] &=&  \frac{k}{\pi} \int d^2z \Big(\p\phi\pb\phi +
(\p\ub+\ub\p\phi)(\pb\u+\u\pb\phi) \Big) +   \nonumber \\
& & \frac{k}{\pi} \int d^2 z \; \p\rho\pb\rho +
\int d^2 z (b\pb c + {\bar b}\p {\bar c}).
\end{eqnarray}
Note that the fields $\phi,\u,\ub$ and $\rho$ have non-trivial
holonomies. In order to perform the path integral we will
decompose them in a periodic part and a holonomy part:
\begin{eqnarray}
\phi & = & {\hat \phi} + \frac{1}{4\tau_2}
\Big((u_1 {\bar \tau} - u_2) z + (u_1 \tau - u_2) {\bar z})\Big)
\nonumber \\
\u & = & {\hat \u} \exp\Big( -  \frac{1}{4\tau_2}
((u_1 {\bar \tau} - u_2) z - (u_1 \tau - u_2) {\bar z}))\Big)
\nonumber \\
\ub & = & {\hat \ub} \exp\Big( + \frac{1}{4\tau_2}
((u_1 {\bar \tau} - u_2) z - (u_1 \tau - u_2) {\bar z}))\Big)
\nonumber \\
\rho & = & {\hat \rho} + \frac{1}{4\tau_2}
\Big((u_1 {\bar \tau} - u_2) z + (u_1 \tau - u_2) {\bar z})\Big),
\label{eq:split}
\end{eqnarray}
where the hatted fields are periodic.\footnote{Trying to follow the 
holonomies of the
gauge field through the field redefinitions we gave before
gives rise to the difficulties we mentioned related to analytic 
continuation and reality of the fields.
We chose the holonomies to be consistent with complex conjugation for $v$,
reality for $\phi$, etc. We believe the resulting spectrum gives sufficient
justification for this choice of analytic continuation.}
%
The coset partition function then reads
\beq
Z_{cs}(\tau) =
\int  {\cal D}{\hat \phi} {\cal D}{\hat\u} {\cal D}{\hat \ub}
{\cal D}{\hat \rho} {\cal D}b
{\cal D}c
\int_{-\infty}^{+\infty} d u_1 d u_2
e^{- S[\phi,\u,\ub;\rho;b,c]} .
\label{eq:pf1}
\eeq
\subsection{Ray-Singer torsion}
The core of the computation uses the Ray-Singer analytic torsion
\cite{RS}, which arises from the
path integral over ${\hat \u}, {\hat \ub}$.
The relevant
piece of the action, after substituting  (\ref{eq:split}), is
\begin{eqnarray}
S_{\u,\ub} &=&\Big(\p + \p{\hat\phi} + \frac{1}{2\tau_2}(u_1 {\bar \tau} - 
u_2) \Big)
\;{\hat \ub} \;
\Big(\pb+ \pb{\hat\phi} + \frac{1}{2\tau_2}(u_1 \tau - u_2) \Big)
\;{\hat \u}
\end{eqnarray}
Note that the action is  quadratic in ${\hat \u}, {\hat \ub}$.
Following \cite{Gawedzki:1991yu}, we
observe that we can disentangle the ${\hat \phi}$-dependence by a chiral
rotation.
The integral over $\u, \ub$ then becomes
the regularised determinant of the Laplacian on a space of
functions that have
non-trivial holonomies around the cycles of the two-torus. Precisely
this determinant was defined in \cite{RS} by using $\zeta$-function
regularisation. The regularised determinant is called the analytic
torsion\footnote{Note that the computation of the analytic torsion
on the torus (cf. \cite{RS}, p. 165-169), naturally resembles the usual
computation of the partition function for a compact boson.}:
\begin{eqnarray}
& {\rm det} \Big|\p  + \frac{1}{2 \tau_2}(u_1 {\bar \tau} -u_2) \Big|^{-2}  =
& \nonumber \\
& \frac{ (q{\bar q})^{-2/24}}{|{\rm sin}(\pi (u_1 \tau - u_2) )|^2}
\frac{e^{\frac{2\pi}{ \tau_2} ({\rm Im} (u_1  \tau - u_2))^2}}{
|\prod_{r=1}^{\infty} (1-e^{2\pi i r \tau - 2\pi i(u_1 \tau - u_2)})
(1-e^{2\pi i r \tau + 2\pi i(u_1  \tau - u_2)})|^2}  &.
\end{eqnarray}
We introduced the
usual notation $q={\rm exp}(2\pi i \tau)$. The analytic torsion
is periodic in the holonomies $u_1$ and $u_2$, as we would expect from
gauge invariance.\footnote{It is also evident from the mathematical
definition of analytic torsion in terms of a complex line bundle with
non-trivial
character $\chi(m \tau + n) = e^{2 \pi i (m u_1 +n u_2)}$. Note that
the authors of \cite{Maldacena:2000kv} appropriately
use an analytically continued version of the
analytic torsion that is not periodic.} If needed (for instance in order to
check  modular properties \cite{RS}), the analytic torsion
can be re-written in terms of the $\theta_1$-function.
\subsection{Free contributions}
In this subsection we treat the other contributions to the partition
function which are basically the familiar free contributions, but
some factors need to be treated with care.
First of all note that there is a shift
$k \rightarrow k-2$ in the kinetic term of
${\hat \phi}$ because of the contribution of the chiral rotation that we
performed to disentangle $\phi$ and $\u, \ub$.
The path integration over
${\hat \phi}$ and ${\hat\rho}$ will each
give the usual periodic boson partition sum
$\tau_2^{-1/2} |\eta(\tau)|^{-2}$ with overall factor $2 \sqrt{k(k-2)}$.
Moreover, the holonomy contributes an overall exponential factor.
Finally, the contribution from
the ghosts $b,c$ that we introduced to gauge fix the $U(1)$ symmetry
is $\tau_2 |\eta(\tau)|^{4}$ \cite{Pol}. It is natural that the net
effect of the gauge field is to cancel the free boson contribution
to the $SL(2,R)$ partition function. \label{free}
\subsection{Holonomies}
It is convenient at this point to break the holonomy parameters
$u_1$ and $u_2$ in integer and fractional parts, i.e. $u_1 = s_1 + w,
\; u_2 = s_2 + m$ with $s_1, s_2 \in [0,1)$ and $w, m \in \mathbb{Z}$
running over the integers. Since the Ray-Singer torsion is periodic,
it is only the overall exponential factor that depends on the integers
$w$ and $m$ that parametrize the non-trivial windings for
the compact bosons.
\subsection{Combining ingredients}
Combining all of the above we obtain for the modular
invariant partition function:
\begin{eqnarray}
Z_{cs}(\tau) &=& 2(k(k-2))^{1/2}
\int_{0}^{1} d s_1 d s_2 \nonumber \\
& & \sum_{w,m = -\infty}^{+\infty}
\frac{ (q{\bar q})^{-2/24} }{ |{\rm sin}(\pi (s_1 \tau - s_2) )|^2 }
\nonumber \\
& &
\frac{  e^{-\frac{k\pi}{\tau_2}|(s_1 + w)\tau-(s_2 + m)|^2 +
\frac{2\pi}{ \tau_2} ({\rm Im} (s_1  \tau - s_2))^2}}{|\prod_{r=1}^{\infty}
(1-e^{2\pi i r \tau - 2\pi i(s_1 \tau - s_2)})
(1-e^{2\pi i r \tau + 2\pi i(s_1  \tau - s_2)})|^2}.
\label{eq:zcs}
\end{eqnarray}
If we are interested in incorporating the coset theory as a factor
in a string theory background $SL(2,R)/U(1) \times {\cal M}$,
we combine it with the modular invariant
partition function
$Z_{\cal M}$ for strings propagating on
${\cal M}$ and the reparametrization ghosts
partition function $Z_{ghosts}$. We then integrate
the modular parameter $\tau$
over the fundamental domain $F_0$ of the usual $SL(2,Z)$ action on
the complex $\tau$-plane to obtain:
\beq
{\cal Z} = \int_{F_0} \frac{d\tau d{\bar \tau}}{\tau_2}  Z_{\cal M}(\tau)
Z_{cs}(\tau) Z_{ghosts}(\tau).
\eeq
The general form of the partition function corresponding
to the background ${\cal M}$ is
\beq
Z_{\cal M}(\tau) = (q{\bar q})^{-c_{\cal M}/24} \sum_{i} q^{h_{i}}
{\bar q}^{{\bar h}_i}
\eeq
where $i$ labels all  states of the CFT on ${\cal M}$  and
$h_{i}, {\bar h}_i$ are the left-moving and
right-moving  conformal weights. Modular invariance
implies that $h_{i} - {\bar h}_i$  is an integer.
By $c_{\cal M}$
we denote the central charge of the  CFT associated to ${\cal M}$.
The total partition function can then be written as:
\newpage
\begin{eqnarray}
{\cal Z} &=& 2(k(k-2))^{1/2}
\int_{F_0} \frac{d\tau d{\bar \tau}}{\tau_2}
\int_{0}^{1} d s_1 d s_2
\nonumber \\
& &
\sum_{w,m = -\infty}^{+\infty}
\sum_{i} q^{h_{i}} {\bar q}^{{\bar h}_i}
e^{4\pi\tau_2(1-\frac{1}{4(k-2)})
-\frac{k\pi}{\tau_2}|(s_1 + w)\tau-(s_2 + m)|^2 +
2\pi\tau_2 s_1^2}  \nonumber \\
& & \frac{1}{|{\rm sin}(\pi (s_1 \tau - s_2) )|^2}
\Bigg|
\prod_{r=1}^{\infty}
\frac{(1-e^{2\pi i r \tau})^2}
{(1-e^{2\pi i r \tau - 2\pi i(s_1 \tau - s_2)})
(1-e^{2\pi i r \tau + 2\pi i(s_1  \tau - s_2)})}
\Bigg|^2.
\label{eq:master}
\end{eqnarray}
Now we need to disentangle the information hidden in this complicated
formula.
\section{Decomposition in characters}
\label{analysis}
We want to connect our partition function computation to
expectations from an algebraic analysis
for the Hilbert space of the coset theory. To that end 
we need to manipulate our result further and determine the character
contributions of the different affine
representations to the partition function. In other words, we have
 to find the correct Hilbert space to trace over that will reproduce
the above partition function. It is appropriate then to first recall
some $SL(2,R)$ representation theory.
See e.g. \cite{Dixon:1989cg} for a more complete treatment.
The representations of the affine
algebras are the modules built on the $SL(2,R)$ representations
using the creation modes of the currents.
The  $SL(2,R)$ representations we will encounter are the (principal)
discrete representations
with lowest weight $D^+_j = \{ |j,m \rangle : m=j, j+1, j+2, \dots \}$ where
the lowest weight state has $J^3_0$ eigenvalue $j>0$ and is annihilated by
$J^-_0$, and similarly for the discrete highest weight representations
$D^-_j = \{ |j,m \rangle : m=j, j-1, j-2, \dots \}$. 
The continuous representations 
$C^{\alpha}_j = \{ |j,m \rangle: m=\alpha, 
\alpha \pm 1, \dots \}$ where $\alpha \in [0,1)$,
have an unbounded $J^3_0$ spectrum 
and $j=\frac{1}{2}+is$ with $s$ real. 
The quadratic Casimir of all these representations is $c_2 = - j (j-1)$.

After refreshing our memory on $SL(2,R)$ representations, we return to
decompose the partition function into a sum over representations.
 We will do this in several steps. We
first write the compact boson part in a more recognizable
form. Secondly, we expand the partition function into a sum over states.
And thirdly we identify contributions from discrete and continuous
representations of $SL(2,R)$. 

 To work towards the
spectrum predicted in  \cite{Dijkgraaf:1991ba}, we first identify
the momentum of the compact scalar. The relevant Poisson resummation is:
\begin{eqnarray}
\sum_{m = -\infty}^{+\infty} e^{-\frac{k\pi}{\tau_2}(m^2 - 2m ((s_1+w)\tau_1
-s_2))} = \sqrt{\frac{\tau_2}{k}} \sum_{n = -\infty}^{+\infty}
e^{-\frac{\pi\tau_2}{k}(n + \frac{i k}{\tau_2}((s_1+w)\tau_1-s_2))^2}
\end{eqnarray}
where we have resummed over $m$ and the new integer
$n \in \mathbb{Z}$ is the momentum of the scalar. Secondly, 
after the Poisson
resummation, we expand the infinite products as well as the sin-prefactor
in (\ref{eq:master})
into an infinite sum of exponential terms.
For a state in the $SL(2,R)$ CFT with levels $N, {\bar N}$
and conformal weights $h,{\bar h}$ in the CFT on ${\cal M}$ (including
reparametrization ghost contributions), the
exponent arising from this expansion is:
\begin{eqnarray}
\mbox{exponent}_{expansion} &=&
2\pi i \tau_1 (N+h - {\bar N} -  {\bar h} + (q-{\bar q})s_1) \nonumber \\
&- & 2\pi \tau_2 (N+h + {\bar N} + {\bar h} + (q+{\bar q} + 1)s_1) -
2\pi i s_2 (q-{\bar q})  \label{expansion}
\end{eqnarray}
where $q$ counts the number of $J^{+}_{n \le 0}$
minus the number of  $J^{-}_{n < 0}$ operators,
corresponding to the particular state under examination. A similar
definition holds for ${\bar q}$ in terms of the right-moving creation
operators.
The overall contribution to the exponent is:
\begin{eqnarray}
\mbox{exponent}_{overall} &=& 4\pi\tau_2(1-\frac{1}{4(k-2)}) +
2\pi i n s_2 -\frac{\pi\tau_2}{k}n^2 -2\pi i n \tau_1 (w+s_1)
\nonumber \\
&& + (2-k)\pi\tau_2s_1^2 - 2 k \pi \tau_2s_1 w - k \pi \tau_2 w^2.
\label{overall}
\end{eqnarray}
Integrating over $s_2$ (see (\ref{expansion}) and (\ref{overall}))
results in
the constraint $q-{\bar q} =  n $. After substituting $q-\bar{q}=n$,
 we find the total
exponent
\begin{eqnarray}
\mbox{exponent}_{total} &=&
2\pi i \tau_1 \Big(N+h - {\bar N} -  {\bar h} - n w\Big)
\nonumber \\
&&- 2\pi \tau_2 \Big(N+h + {\bar N} + {\bar h} + (q+{\bar q} + 1)s_1
\nonumber \\
&& -2 (1-\frac{1}{4(k-2)})
+ \frac{n^2}{2k} + \frac{k}{2}w^2 + kws_1 + \frac{k-2}{2}s_1^2
\Big)
\end{eqnarray}
The integral over the first holonomy was fairly easy, and gave us one
of the expected constraints \cite{Dijkgraaf:1991ba}. It relates
the momentum of the compact boson to the $J^3_0-\bar{J}^3_0$ eigenvalue
in the $SL(2,R)$ representation.

The integral
over the second holonomy is far less trivial and needs some technical
trickery, inspired by the analysis in \cite{Maldacena:2001km}. It will
allow us to separate the contributions from discrete and continuous
representations of $SL(2,R)$.
We first introduce an auxiliary variable to incorporate a prefactor
and the piece of the exponent quadratic in $s_1$:
\begin{eqnarray}
\sqrt{(k-2) \tau_2} e^{-2\pi\tau_2(\frac{k-2}{2} s_1^2 + (k w + 1
+ (q+{\bar q}))s_1)} =
\int_{-\infty}^{+\infty} dc \; e^{-\frac{\pi}{(k-2)\tau_2} c^2
-2 \pi (i c + \tau_2 (k w + 1 + (q+{\bar q}))) s_1}. \nonumber
\label{eq:aux}
\end{eqnarray}
The integration over $s_1$ is now straightforward:
\begin{eqnarray}
&\int_{0}^{1} ds_1 e^{-2\pi s_1 (i c + \tau_2 (kw + 1 + (q + {\bar q}))
)} = & \nonumber \\
&  \frac{-1}{2\pi(i c + \tau_2 (kw + 1 + (q + {\bar q})))}
\Big( e^{-2\pi  (i c + \tau_2 (kw + 1 + (q + {\bar q})))} - 1 \Big) &
\end{eqnarray}
Combining it with the quadratic term in $c$
results in the term
\begin{eqnarray}
\frac{-1}{2\pi(i c + \tau_2 (kw + 1 + (q + {\bar q})))}
\Big(
e^{-\frac{\pi}{(k-2)\tau_2} c^2
-2\pi  (i c + \tau_2 (kw + 1 + (q + {\bar q})))} -
e^{-\frac{\pi}{(k-2)\tau_2} c^2}
\Big).
\label{eq:c-terms}
\end{eqnarray}
\subsection{Discrete representations}
Now we observe that the exponent of the first term can be completed
to a square if we set $c =2 \tau_2 s - i \tau_2 (k-2)$. Shifting the contour
of $c$ (for the first term only)
from ${\rm Im} \, c = 0$ to $ {\rm Im} \, c = -i \tau_2 (k-2)$,
picks up residues from the poles of the denominator
in the range $ -\tau_2(k-2) < {\rm Im} \, c < 0$. The poles are located
at $c = i \tau_2 (kw + 1 + (q+{\bar q}))$ in the range:
\begin{eqnarray}
-\tau_2(k-2) < \tau_2 (kw + 1 + (q+{\bar q})) < 0.
\label{poleregion}
\end{eqnarray}
Now we note that we
 can interpret the pole contributions to the integral summed
over $q, \bar{q}, w, n$, as the trace over a constrained
Hilbert space. Consider
the product Hilbert space $\hat{{\cal D}}^+_j$, the module built on
the discrete representation ${\cal D}^+_j$ of $SL(2,R)$, and the Hilbert
space for the compact  boson ${\cal H}^{U(1)}$.  The first constraint
we put on the sum over states is $J^3_0-\bar{J}^3_0=n$, namely the
constraint we obtained from the $s_2$-integration. The second constraint
determines the quadratic Casimir $j$ of the $SL(2,R)$ representation
in terms of the winding number of the compact boson:
 $J^3_0+\bar{J}^3_0 = - kw$ or equivalently $kw + 1 +
(q+{\bar q}) = 1-2 j$. One way to see the necessity for this constraint
is the fact that the T-duality $(J^3,{\bar J}^3,n,w) \rightarrow 
(\frac{1}{k}J^3,- \frac{1}{k}{\bar J}^3,-w,-n)$ is a symmetry
of our partition function (which is reflected in the constraint
equations).
 
The discrete nature of the representations
is determined by the fact that the sin-prefactor gives rise to only one
kind of operator at level zero, namely $J^+_0$, and not to
$J^-_0$ contributions.
Notice moreover that we needn't sum over creation
operators for the $J^3$-current or for the compact boson, since
their contributions to the partition function were cancelled by
the $U(1)$ ghosts. The second constraint $kw + 1 + (q+{\bar q}) =
1-2 j$, immediately implies (via \ref{poleregion})
 the expected bounds on $j$, the Casimir
of the discrete representation:
\begin{eqnarray}
\frac{1}{2} < j <\frac{k-1}{2}. \label{bound}
\end{eqnarray}
We emphasize that the upper bound we derived is not the one suggested
in \cite{Dijkgraaf:1991ba} but the improved
 bound\footnote{The improved bound was suggested for the ungauged model on the
basis of consistency with the inclusion of spectral flowed representations
in \cite{Maldacena:2000hw} and on the basis of fusion rules in 
\cite{Gawedzki:1991yu}. The improved bound was shown 
to be necessary in the coset model for a tachyon free spectrum in Little 
String Theory in
\cite{Giveon:1999px}. We prove this consistency 
requirement.} derived in
\cite{Maldacena:2000kv} for the ungauged
$SL(2,R)$ WZW model.\footnote{As we will 
see in the following, a
continuous spectrum opens up when  $j$ reaches either
the lower or the upper
bound \cite{Maldacena:1998uz,Seiberg:1999xz}.} Using the
constraint we can rewrite the exponent in a familiar form. We
obtain a sum over the described Hilbert space ${\rm Tr}_{\hat{{\cal
D}}^+_j \otimes \, \hat{{\cal
D}}^+_j}  q^{L_0^{cs}}
\bar{q}^{\bar{L}_0^{cs}}$ where the $L_0^{cs}$ operator takes the
standard form:
\begin{eqnarray}
L_0^{cs} &=& L_0^{SL(2,R)} - L_0^{U(1)}.
\end{eqnarray}
The conformal weights of the primary states, which agree
with the total exponent after substitution of the values
for the poles, are given by:
\begin{eqnarray}
h_{cs} &= &  -\frac{j(j-1)}{k-2} + \frac{(n-kw)^2}{4k}   \\
{\bar h}_{cs} & = &  -\frac{j(j-1)}{k-2} + \frac{(n+kw)^2}{4k}.
\end{eqnarray}
The summation is  over states
with the constraints $J^3_0-\bar{J}^3_0 = n $, $J^3_0+\bar{J}^3_0 = -kw$
and no contribution from the $J^3_{n < 0}$ oscillators.
Thus we interpreted the first part of our partition function as a
character over a constrained product of an affine discrete $SL(2,R)$
representation
times a compact boson. We sum over discrete representations
that satisfy the bound (\ref{bound}). For the parafermion
interpretation of this Hilbert space we refer to
\cite{Lykken:1988ut,Maldacena:2000hw}. 

We remark that we crucially made use of the periodicity of the Ray-Singer
torsion in the $u_1$-variable in our computation. If we would ignore
this periodicity, it is clear from
the analysis in \cite{Maldacena:2000kv} that we could
identify the winding number $w$
of the compact boson with the parameter $w$ that controls the
expansion of the different products in the denominator of the partition
function, and therefore with the spectral flow parameter in the
$SL(2,R)$-WZW model.\footnote{Note that this also follows from
the fact that the integer holonomies $w$ change the current algebra 
on the torus according to spectral flow. 
See e.g. \cite{Kiritsis:1993ju}.} This exemplifies
in detail the relation uncovered in  \cite{Maldacena:2000hw} between
spectral flow and the winding of strings,  in the coset
model.
\subsection{Continuous representations}
We combine now the shifted integral over $s$ of the first term in 
(\ref{eq:c-terms}) with the integral over the second term, in which we 
rescale
$c=2\tau_2 s$. Including the summation over winding numbers $w$, we obtain
\begin{eqnarray}
&-\frac{1}{\pi} \sum_{w = -\infty}^{+\infty} \int_{-\infty}^{+\infty} ds
\Bigg[\frac{e^{-2\pi\tau_2
(N + h + {\bar N} + {\bar h} - 2 + 2 \frac{s^2+1/4}
{k-2} + \frac{n^2}{2k} + \frac{k}{2} (w+1)^2 + (q+{\bar q}))}}
{2 i s + k (w+1) -1 + (q+{\bar q})}&\nonumber\\
&
- \frac{e^{-2\pi\tau_2
(N + h + {\bar N} + {\bar h} - 2 + 2 \frac{s^2+1/4}
{k-2} + \frac{n^2}{2k} + \frac{k}{2} w^2)}}
{2 i s + k w + 1 + (q+{\bar q})} \Bigg] & .
\end{eqnarray}
Inspecting
the above expression, we observe that the first term of the $w-1$ sector
and the second term of the $w$ winding sector share the same exponent after
spectral flow of the first by one unit
$N, {\bar N} \rightarrow N + q, {\bar N} + {\bar q}$. The last operation
is based on the isomorphism
$\hat{{\cal D}}^{+,w-1}_j \cong \hat{{\cal D}}^{-,w}_{\frac{k}{2}-j}$
where the second upper index denotes the amount of spectral 
flow \cite{Maldacena:2000hw}, i.e. these are discrete representations
defined with respect to the algebra obtained after spectral
flow.
  Combining
terms this way we get
\begin{eqnarray}
&-\frac{1}{\pi} \sum_{w = -\infty}^{+\infty} \int_{-\infty}^{+\infty} ds \;
e^{-2\pi\tau_2
(N + h + {\bar N} + {\bar h} - 2 + 2 \frac{s^2+1/4}
{k-2} + \frac{n^2}{2k} + \frac{k}{2} w^2)}
&\nonumber\\
&\Big(\frac{1}{2 i s + k w -1 + (q+{\bar q})}
-\frac{1}{2 i s + k w + 1 + (q+{\bar q})}\Big)&
\end{eqnarray}

As in  \cite{Maldacena:2000kv}, these two terms
can be interpreted as representing two halves of a continuous representation
with $j=\frac{1}{2} + i s$. The first term represents the contribution
of a $D^-$ representation (after spectral flow) and the second term still
corresponds to a $D^+$ representation. In particular, note that the second
term,
when summed over states $(J^+_0 \bar{J}^+_0)^r |\psi\rangle$ in ${\cal 
D}^+$,
gives rise to a logarithmically divergent sum. We adopt here
the regularisation procedure  of \cite{Maldacena:2000kv} and introduce
a Liouville wall that cuts off the infinite volume otherwise available
to the strings in the continuous representation\footnote{A rigorous 
justification of this 
procedure would require a precise identification of
the coefficient of the exponential
suppression after the introduction of a Liouville wall at a 
finite distance
in the target space \cite{Maldacena:2000kv}, and  a precise
treatment of the sum over the $J^3_0$ charge that is related to the
creation and annihilation operators of the $J^+$ and $J^-$ currents.
As in \cite{Maldacena:2000kv}, we will find justification for the adopted
prescription from an independent scattering amplitude argument.
}. Thus we obtain
a regularised sum over the zeromodes of the following form:
\begin{eqnarray}
-\frac{1}{2} \sum_{r = 0}^{\infty} \frac{1}{A+r} e^{-r\epsilon} = 
-\frac{1}{2}\log \epsilon + \frac{1}{2}\frac{d}{dA}\log\Gamma(A),
\;\;\;A = i s + \frac{1}{2} (kw + 1 - n).
\end{eqnarray}
Similarly, the first term in the integral can be interpreted as an infinite 
sum
over states in a
${\cal D}^-$ Hilbert space of the form $(J^-_0 \bar{J}^-_0)^r|\psi\rangle$
\begin{eqnarray}
\frac{1}{2} \sum_{r = 0}^{\infty} \frac{1}{B-r} e^{-r\epsilon}= 
-\frac{1}{2}\log \epsilon- \frac{1}{2}
\frac{d}{dB} \log \Gamma(-B), \;\;\;
B = i s + \frac{1}{2} (kw - 1 + n).
\end{eqnarray}
The density of states as a function of $s$ 
is then found to be
\begin{eqnarray}
\rho(s) =  \frac{1}{2\pi} 2\log\epsilon + \frac{1}{2\pi i} \frac{d}{2 ds} \log
\frac{
\Gamma(-i s + \frac{1}{2} - m)\;\Gamma(-i s + \frac{1}{2} + {\bar m})}
{\Gamma(+ i s + \frac{1}{2}+{\bar m})\;\Gamma( +is + \frac{1}{2} - m)}
, \label{density}
\end{eqnarray}
where $m= \frac{1}{2}(n-kw), \; {\bar m} = -\frac{1}{2}(kw + n)$ are the
eigenvalues of $J^3_0$ and $\bar{J}^3_0$.
In the above expression we have truncated the range of integration over $s$
to $[0,\infty)$
using the invariance of the exponent under $s \rightarrow -s$. Thus,
the contribution
of the continuous representations of $SL(2,R)$ combined with the
momentum and winding modes of the free boson, can be written as
\begin{eqnarray}
\sum_{w, n = -\infty}^{+\infty} \int_{0}^{+\infty} 2ds \;\rho(s)
{\rm Tr}_{\hat{{\cal C}}_{\frac{1}{2}+is} \otimes \, \hat{{\cal C}}_{\frac{1}{2}+is}} q^{L^{cs}_0} {\bar 
q}^{{\bar L}^{cs}_{0}}
\end{eqnarray}
where the conformal primaries have weights
\begin{eqnarray}
h_{cs} &=& \frac{s^2 + \frac{1}{4}}{k-2} +  \frac{(n-kw)^2}{4k} \\
{\bar h}_{cs} &=& \frac{s^2 + \frac{1}{4}}{k-2} +  \frac{(n+kw)^2}{4k}
\end{eqnarray}
and the trace over $\hat{C}_{\frac{1}{2}+is} \otimes \, \hat{{\cal C}}_{\frac{1}{2}+is}$ 
is subject to the same constraints as before, namely $J^3_0+\bar{J}^3_0= - kw$,
$J^3_0-\bar{J}^3_0=n$ and the $J^3$-current and free boson creation
operators act trivially.

As in \cite{Maldacena:2000kv}, we can perform a consistency check on the
density of states by
analysing the phase shift in a scattering experiment.
We can introduce a Liouville wall for the continuous representation
strings, to cut off
the infinite volume available to them\footnote{In our case the volume
divergence is apparent from the pole of the partition function 
(\ref{eq:master})
at $s_1=0=s_2$. Excising the pole corresponds to introducing the Liouville
wall.}, 
and relate the density of states to
the phase shift for
scattering a string in the bulk of $SL(2,R)/U(1)$ and then off the 
Liouville wall \cite{Maldacena:2000kv}.
Making use of the fact that
the form of the scattering amplitude is the same
for the coset theory as for the ungauged $SL(2,R)$ model
(see e.g. \cite{Giveon:1999px}), we can conclude that the density
of states is indeed given by (\ref{density}), where we obtain the 
eigenvalues of the $J^3_0$ and $\bar{J}^3_0$ operators from the constraint
equations on the Hilbert space. This gives an overall consistency check
on our regularisation procedure.

\section{Conclusions}
\label{conclusions}
We have thereby finished the identification of the characters of the 
different
representations in the $SL(2,R)/U(1)$ coset partition function. We obtained
agreement with the spectrum that one would get by imposing the usual
\cite{Dijkgraaf:1991ba}
algebraic constraints on the spectrum derived in \cite{Maldacena:2000hw}
for the ungauged model. We have thereby proved
the correctness of this procedure from first principles. 
In particular we proved the upper bound on the spin
and determined the density of string states in the continuous
representations. 

It is straightforward to extend our computation
to the T-dual trumpet background 
\cite{Giveon:1991sy, Dijkgraaf:1991ba}, since our
analytic treatment is related smoothly  to the algebraic treatment of
the conformal field theory in which T-duality is manifest. The
partition function
of the  Lorentzian black hole too, should now be within reach.
It might also be possible to include a mass in the black hole background
\cite{Witten:1991yr} and compute the temperature
dependent partition function. This could provide a setup
where aspects of stringy black hole thermodynamics could be addressed
in a systematic manner. 

The program of studying the coset theory can of
course be followed along the lines of \cite{Maldacena:2001km} by
computing various two-, three- and four-point correlation functions to check
the completeness of the Hilbert space. This should be straightforward
due to the easy relations between the correlations functions for the
parent and the coset theories.
Note  that we can interpret the winding number
$w$ and the momentum $n$ as the winding of the string around the semi-infinite
cigar, and the momentum around the circle at infinity. The winding number
 is not
conserved since the string can slip off the semi-infinite cigar, but it
is well possible that a precise 
restriction on the violation of winding number can be
derived following  \cite{Maldacena:2001km} (appendix D).

One of our prime motivations for computing the partition function on
the black hole background was to be able to analyse more rigorously
the spectrum of string theory on backgrounds including the
$SL(2,R)/U(1)$ coset. Given our analysis this is now possible and should
find applications, for example
in analysing the holographic correspondence for these
backgrounds.

It is also important to construct boundary states in the
coset theory that would correspond to D-branes in these backgrounds.
For the parent $SL(2,R)$ conformal field theory, the construction was
done in 
\cite{Giveon:2001uq,Rajaraman:2001cr,Parnachev:2001gw,Ponsot:2001gt,Lee:2001xe,Lee:2001gh},
with applications to $AdS_3$ string theory
\cite{Bachas:2000fr,Bachas:2001vj,Karch:2001cw,DeWolfe:2001pq}. 
For the coset theory a similar
construction should be possible 
\cite{Yamaguchi:2001rn, Ponsot:2001gt}, and should yield
information on D-branes in NS5-brane backgrounds (see e.g.
\cite{Elitzur:2000pq,Gava:2001gv}).

Another extension of our results would be
to analyse the partition function of the $N=2$
supersymmetric extension of our coset model and make connection with the
work \cite{Hori:2001ax} (based on \cite{FZZ}).
Namely, a precise analysis of the spectrum
 might shed additional light on the duality
between the 
$SL(2,R)/U(1)$ coset theory and the $N=2$ Liouville CFT.

In summary, we believe that we provided a good basis for a precise path 
integral
analysis of the spectrum of non-compact coset conformal field theories and for
uncovering more secrets of black hole backgrounds in CFT and string theory.

\acknowledgments
Research supported in part by the Reed Fund Award,
the CTP and the LNS of MIT and the U.S. Department of Energy
under cooperative research agreement \# DE-FC02-94ER40818.
A. H. is also supported by an A. P. Sloan Foundation Fellowship,
and a DOE OJI award.

\end{document}